\documentclass[journal=jacsat,manuscript=article]{achemso}
\usepackage{chemformula}
\usepackage[T1]{fontenc}
\usepackage{verbatim}
\usepackage{upgreek}
\usepackage{lineno}

\author{Nadine Denis}
\affiliation[University of Basel]{Physics Department, University of Basel, Basel, Switzerland}

\author{Didem Dede}
\affiliation[EPFL]{Laboratory of Semiconductor Materials, Institute of Materials, EPFL, Lausanne Switzerland}

\author{Timur Nurmamytov}
\affiliation[CNR]{CNR - Istituto Officina dei Materiali (IOM), Laboratorio TASC, Trieste, Italy}
\alsoaffiliation[University of Trieste]{Physics Department, University of Trieste, Trieste, Italy}

\author{Salvatore Cianci}
\affiliation[Sapienza Università di Roma]{Physics Department, Sapienza Università di Roma, Rome, Italy}
 
\author{Francesca Santangeli}
\affiliation[Sapienza Università di Roma]{Physics Department, Sapienza Università di Roma, Rome, Italy}

\author{Marco Felici}
\affiliation[Sapienza Università di Roma]{Physics Department, Sapienza Università di Roma, Rome, Italy}

\author{Victor Boureau}
\affiliation[EPFL]{Interdisciplinary Center for Electron Microscopy, EPFL, Lausanne, Switzerland}

\author{Antonio Polimeni}
\affiliation[Sapienza Università di Roma]{Physics Department, Sapienza Università di Roma, Rome, Italy}

\author{Silvia Rubini}
\affiliation[CNR]{CNR - Istituto Officina dei Materiali (IOM), Laboratorio TASC, Trieste, Italy}

\author{Anna Fontcuberta i Morral}
\affiliation[EPFL]{Laboratory of Semiconductor Materials, Institute of Materials, EPFL, Lausanne, Switzerland}
\alsoaffiliation[EPFL]{Faculty of Basic Sciences, Institute of Physics, EPFL, Lausanne, Switzerland}
	
\author{Marta De Luca}
\email{marta.deluca@uniroma1.it}
\affiliation[Sapienza Università di Roma]{Physics Department, Sapienza Università di Roma, Rome, Italy}

\title {Single photon emitters in thin GaAsN nanowire tubes grown on Si}

\keywords{Dilute nitrides, GaAsN, nanowires, single photon emitters, photoluminescence}

\begin{document}

\begin{abstract}
III-V nanowire heterostructures can act as sources of single and entangled photons and are enabling technologies for on-chip applications in future quantum photonic devices. The unique geometry of nanowires allows to integrate lattice-mismatched components beyond the limits of planar epilayers and to create radially and axially confined quantum structures. Here, we report the plasma-assisted molecular beam epitaxy growth of thin GaAs/GaAsN/GaAs core-multishell nanowires monolithically integrated on Si (111) substrates, overcoming the challenges caused by the low solubility of N and a high lattice mismatch. The nanowires have a GaAsN shell of 10 nm containing 2.7\% N, which reduces the GaAs bandgap drastically by 400 meV. They have a symmetric core-shell structure with sharp boundaries and a defect-free zincblende phase. The high structural quality reflects in their excellent opto-electroinic properties, including remarkable single photon emission from quantum confined states in the thin GaAsN shell with a second-order autocorrelation function at zero time delay as low as 0.056.

\end{abstract}

\section{Introduction}
The unique growth, transport, and optical properties of semiconductor nanowires (NWs) have attracted significant research interest, making them valuable components for nanophotonic and quantum optical devices. NWs have emerged as a platform for applications such as subwavelength lasers and ultrasensitive sensors \cite{Eaton2016}, and may be integrated into quantum communication or quantum computation devices as sources of single photons \cite{https://doi.org/10.1002/adfm.202315936, Haffouz2018, Mäntynen2019} and entangled photon pairs \cite{Khoshnegar2017}. Furthermore, NW-based devices can be integrated on-chip into an optical cavity or photonic crystal by growing them site-controlled or by individually moving them to a desired location \cite{Mäntynen2019, Takiguchi2020}. 

With NWs, in contrast to thin films, the superior physical properties of III-V materials — such as their high electron mobility and direct bandgap — can be grown more easily bottom-up onto standard Si substrates. This is due to the relaxed lattice mismatch requirements owing to a small interface between the NW and the Si surface \cite{Glas2006, Cirlin2009, Royo2017}. Moreover, NWs can accommodate strains in two dimensions due to the large surface-to-volume ratio, making them suitable for growing core-shell heterostructures with highly lattice-mismatched materials \cite{Glas2015, McIntyre2020, Balaghi2019}. During the NW growth, defects with long-range strain fields such as vacancies and dislocations are typically attracted to nearby surfaces, making them inherently free of point and line defects \cite{Sanchez2017}, which affect carrier mobility and lifetime. Furthermore, in NW shells made of ternary alloys such as the GaAsN-based heterostructures investigated here, strain can be compensated by radial variations in the alloy composition at the corners \cite{Rudolph2013, Balaghi2019, Balagula2020}. For all these reasons, the NW geometry offers an enhanced ability to control the bandgap across radial and axial heterostructures while maintaining the high crystal quality required for nanoscale devices.

This work introduces GaAs/GaAsN NWs to the panorama of III-V single photon emitters that can be monolithically integrated on Si \cite{Arakawa2020}. So far, this was achieved at wavelengths around 760 nm in AlGaAs/GaAs NWs \cite{Leandro2018}, in the near-infrared around 900 nm in InAs/GaAs NWs \cite{Kwoen2013}, and at telecommunication wavelengths around 1340 nm in InAs/InP NWs \cite{Jaffal2019}. Here, we obtain single photon emitters with wavelengths around 950-1000 nm by using GaAs/GaAsN-based NWs. Specifically, we achieve single photon emission from vertical NWs, which makes them especially suitable for fibre-coupled devices.

Heterostructures made of a Ga(In)AsN alloy are used in particular for optical applications in the near-infrared range, as the GaAs bandgap decreases significantly when diluted amounts of N are incorporated. A dilute amount of N in GaAs creates a strong perturbation potential in the GaAs lattice, which leads to a splitting of the otherwise degenerate conduction band and drastically reduces the electronic bandgap. According to the band-anti-crossing model (BAC) \cite{Shan1999, Wu2002, Vurgaftman2003}, the bandgap of GaAsN can be continuously reduced by 440 meV by varying the N concentration from 0 to 4\%, see the calculated nitrogen-dependent bandgap energy in the inset of Figure \ref{fig:growth}(c) (orange line). At the same time, the electron effective mass increases \cite{Masia2006}, which reduces the spill-out of carriers in the shell and improves the thermal stability of the confinement. Interestingly, the bandgap reduction is reversible when the material is exposed to low energy ionized hydrogen gas \cite{Felici2006}. This has been used in planar materials to tailor the electronic band structure and create site-controlled single photon emitters on GaAs substrates \cite{Felici2020, Birindelli2014}.
In general, the growth of high quality dilute GaAsN is associated with considerable difficulties due to the large miscibility gap of the constituents, due to the low solubility of N in GaAs and due to a large lattice mismatch between GaAs and GaAsN. By lowering the growth temperature, thus growing GaAsN under non-equilibrium conditions, these problems may be partially overcome, leading to an increase in the N concentration \cite{Ahn2014, Stringfellow2021}. Nonetheless, by slightly increasing the growth temperature to an intermediate range, better optical qualities and less band-edge fluctuations are observed \cite{Buyanova1999b}. Growing dilute nitride NWs with GaAsN in the core has not yet been achieved because the reduced growth temperature required by GaAsN leads to a reduced Ga mobility, which may prevent the Ga adatoms from diffusing from the NW sidewalls up to the Ga droplet at the tip \cite{Kuang2012, La2017}. Furthermore, in the case of dilute GaPN, it has been argued that the N plasma increases the nucleation rate at the Ga-droplet/Si interface, resulting in planar growth instead of vertical NWs at the start \cite{La2016}. Despite all these difficulties, the growth of GaAs/GaAsN/GaAs with N concentration up to 3\% has been reported in thick core-shell NWs \cite{Yukimune2019}, having a total diameter of 350 nm and a GaAsN shell thickness of 50 nm. Although these NWs exhibited twinning and random switching between wurtzite (WZ) and zincblende (ZB) phases, as well as phase separation for high N concentrations, they showed good optical emission properties at room temperature. Contrarily to GaAsN epilayers, incorporating N did not lead to an optical degradation compared to the GaAs reference NWs. Other studies explored the possibility of a patterned growth, for low N concentrations with 0.8\% N and a diameter of 220 nm \cite{La2017}; Also, lately, NWs with a multi-quantum-well GaAs/GaInAsN structure with optical emission at telecommunication wavelength were demonstrated, by alloying with indium \cite{Nakama2023}. These growth efforts are motivated by the unique properties of dilute Ga(In)AsN, which have led to exceptional results for applications such as lasing \cite{Chen2017} or spin filtering in N-rich nanopillars up to room temperature \cite{Chen2018}. In dilute GaNAsP NWs, it was possible to measure single photon emission, with second-order correlation function at zero delay ($g^{(2)}(0)$) values of 0.45 \cite{Jansson2021} at a wavelength of about 700 nm; however, single photon emission in GaAsN-based NWs has not yet been reported.

In this work, we report the growth of thin GaAs/GaAsN/GaAs NWs on Si(111) substrates with a pure ZB phase along the NW axis and a nitrogen concentration of about 2.7\%. The structural properties of the different NW samples were investigated by imaging with transmission electron microscopy (TEM). All the NWs have a pure defect-free ZB structure with a short WZ segment at the tip. The ultramicrotome cuts reveal a symmetric core-shell structure with sharp boundaries. The high crystalline quality of the NWs leads to $\upmu$-photoluminescence (PL) up to room temperature and to the emission of intense and narrow excitonic lines at low temperature. Finally, we achieve single photon emission by exploiting small variations in the N concentration and by designing the GaAsN active region very thin in order to favor quantum confinement of excitons. We measure a multi-photon emission probability below 6\%.

\section{Results and discussion}
\subsection{Nanowire samples}
\begin{figure} []
	\includegraphics{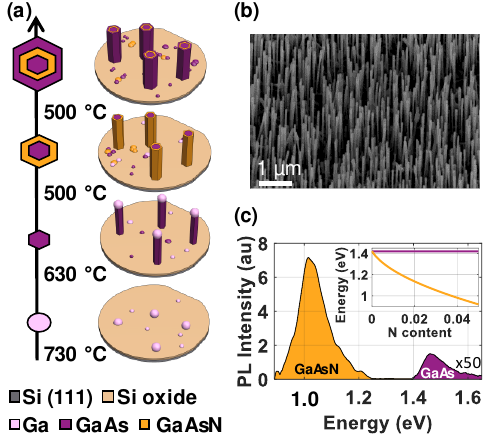} 
	\caption{Growth and design of the GaAs/GaAsN/GaAs core-shell-shell heterostructured NWs. (a) illustrates the NW growth scheme: First, the nucleation of Ga droplets, second, the VLS growth of the thin GaAs core that is terminated with the controlled crystallization of the Ga droplet, third, the VS growth of the thin GaAsN shell at reduced temperature, fourth, the VS growth of the GaAs outer layer. (b) SEM image of the NWs on sample A. (c) $\upmu$-PL spectrum at room temperature showing the bandgap emission of the GaAs-core at 1.46 eV and the emission of the GaAsN-shell at 1.02 eV. The inset shows the reduction of the GaAsN bandgap energy as a function of the N concentration according to the band-anti-crossing model at room temperature for bulk (orange line), while the horizontal purple line marks the GaAs bandgap energy value as a reference.}
	\label{fig:growth} 
\end{figure}
The designed NW heterostructure consists of a GaAs core with higher bandgap energy, surrounded by a thin GaAsN shell with lower bandgap energy and a second high bandgap GaAs outer shell. The NWs are grown by molecular beam epitaxy (MBE) on a Si(111) substrate. A schematic of the NW growth process and resulting geometry is shown in Figure \ref{fig:growth}(a). First, the GaAs core is grown via a Ga-assisted vapor-liquid-solid (VLS) approach at $630^{\circ}$C. Second, the substrate temperature is reduced to $500^{\circ}$C for the epitaxial vapor-solid (VS) growth of the GaAsN and GaAs shells. We have grown three different samples: sample A has a GaAs nominal core diameter of 20 nm, a GaAsN shell of 10 nm, and a GaAs outer shell of 10 nm. Sample B has a thicker core of 40 nm and the same shell thicknesses as sample A. The scanning electron microscopy (SEM) image of the NWs from sample A in Figure \ref{fig:growth}(b) shows that the NWs are uniform, straight, and vertical, with an average length of approximately 2 $\upmu$m. The $\upmu$-PL spectrum of the NW ensemble at room temperature in Figure \ref{fig:growth} (c) reflects the designed heterostructure. The bandgap of the GaAsN-shell is decreased by 400 meV with respect to the emission energy of pure ZB GaAs at 1.42 eV, down to 1.02 eV. The thin GaAs core emits at 1.46 eV. The full width at half maximum (FWHM) of the GaAsN emission is relatively large, with 100 meV compared to typical values in the range of 50 meV in optimized GaAsN epilayers. The broadening might be due to some N concentration or strain fluctuations in the radial direction of the NWs, these are typical in the corners of ternary alloy NW shells \cite{Rudolph2013, Balaghi2019}. According to the BAC model for bulk, we find a N concentration of 3.4\%. However, the true N concentration is expected to be lower, because the smaller lattice constant of GaAsN with respect to GaAs causes a tensile strain in the shell that further reduces the bandgap. This is mirrored in the up-shift of the GaAs emission energy by 40 meV due to compressive strain in the core. Such strain distributions are typical for thin core-shell NW heterostructures with a large lattice mismatch and they have a big impact on the bandgap \cite{Glas2006, Balaghi2019}. For a reduction of the bandgap due to strain in a range of 30-70 meV and considering a blueshift of approximately 25 meV due to quantum confinement in a thin GaAsN quantum well \cite{Wu2002} the N concentration in the shell is between 2.8-3.4\%.

\subsection{Structural characterization}
\begin{figure*}[]
	\centering 
	\includegraphics[width=\textwidth]{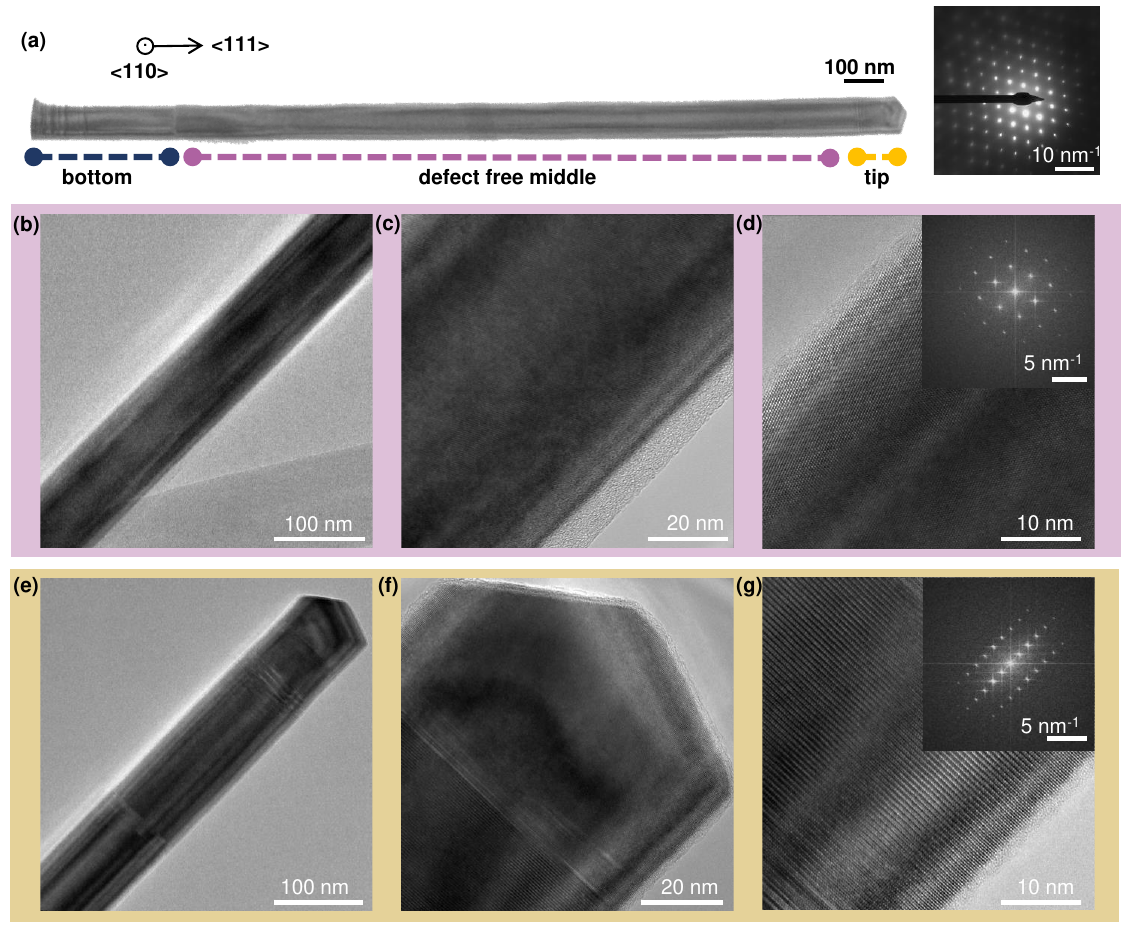} 
	\caption{Structural analysis of a single NW by TEM. (a) BF-TEM image of an entire NW with an inset of the SAED pattern. (b)-(d) HR-TEM images from the middle region with a pure defect-free ZB phase and (e)-(g) HR-TEM images from the tip of the NW with a short WZ segment. The insets in (d) and (g) display the fast Fourier transform showing the ZB phase in (d) and the WZ phase in (g). These images are taken from the <110> zone axis on a NW transferred from sample A.}
	\label{fig:growth1}
\end{figure*}
\begin{figure}[]
	\includegraphics[scale=1]{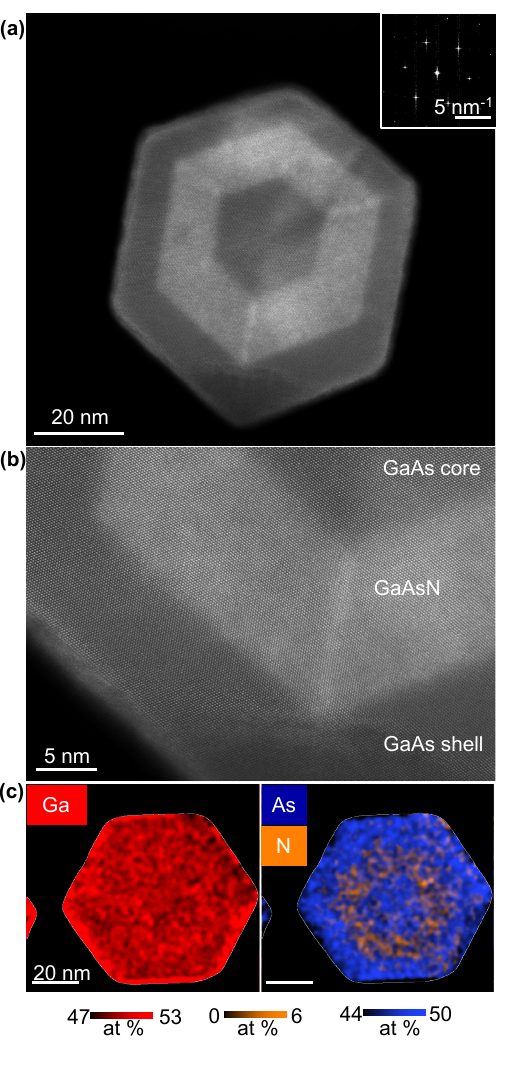} 
	\caption{Structural and compositional analysis of the multi-shell cross-section of sample A. (a)-(b) Atomic-resolution ADF-STEM images from the <111> zone axis. The Fourier transform in the inset of (a) evidences the ZB crystal structure. A brighter contrast is observed in the GaAsN shell along three out of the six {112} planes laying in the symmetry axes linking the corners of the hexagonal section. (c) EDX maps of the atomic concentrations of Ga, and As, N; measuring a N concentration of 2.9\% in the shell.}
	\label{fig:growth2} 
\end{figure}
First, we analyzed the structure of single NWs from Samples A and B by transmission electron microscopy (TEM). Figure \ref{fig:growth1} shows a representative structural characterization of a NW from sample A. Panel (a) shows the bright-field (BF) TEM image of the NW observed in <110> zone axis together with the selected area electron diffraction (SAED) pattern, (b)-(d) show the high resolution (HR) TEM images from the central region of the NW and in (e)-(g) show the HR TEM images from the tip region. HR images from the bottom part of the NW and further characterizations of NWs from sample B are presented in the supporting information (SI1-SI4).
The NWs from sample A and B show a pure and defect-free ZB section and only the bottom and the tip contain twin planes and stacking faults. These are common at the bottom of self-catalyzed NWs as a result of the contact angle instability of the initial droplet \cite{balgarkashi_facet-driven_2020, plissard_gold-free_2010}. It is possible to minimize these by growing at higher temperatures and lower Ga fluxes \cite{krogstrup_structural_2010}. The short WZ segment at the tip is created during the droplet consumption under excess As flux when the contact angle changes. At contact angles between 100 and 125 degrees, the WZ phase is favorable, whereas, at smaller and larger contact angles, the ZB phase is favorable \cite{panciera_phase_2020, Ambrosini_growth_2011}. The high crystal purity and phase stability in the central region are achieved by keeping the diameter of the NW small, such that the droplet is stable over a wide V/III ratio \cite{tersoff_stable_2015}. For thicker core diameters ZB and WZ have a tendency to coexist along the growth direction of the NW \cite{yukimune2018, yukimune2020}. The surrounding GaAsN and GaAs outer shells inherit the crystal structure of the core for these small diameters and N concentrations. We do not observe tapering in the NW diameter along the growth axis. Sample B has similar structural characteristics as sample A.

Next, we investigated the core-shell structure by imaging the axial cross-sectional cuts of the sample prepared by ultramicrotomy. Figures \ref{fig:growth2}(a) and (b) show atomic-resolution images of the sample by annular dark-field (ADF) scanning transmission electron microscopy (STEM). The GaAs/GaAsN/GaAs core-multishell geometry has a well-defined hexagonal shape with \{110\} sidewalls. The GaAs core thickness of NWs from samples A and B is ~23 and ~42 nm, respectively (SI5 shows the precise measurements). The GaAsN shell thickness is $\sim$10 nm in both samples, which is very close to their nominally defined size. A distinct brighter contrast is observed along three out of six {112} planes laying in the symmetry axes linking the corners of the hexagonal shape of the NW. These may be caused by polarity-driven segregation on the <112> directions, which can be either A or B polar, depending on whether they are terminated by group III or group V compounds. Similar phenomena have been observed in various ternary alloy core-shell structures, where variations in the material composition could be measured \cite{Zheng2013, Rudolph2013, Balaghi2019}, see SI6 and SI7 for further measurements and discussions. Finally, the compositions within the NW heterostructure is analysed with energy dispersive x-ray spectroscopy (EDX). Quantitative elemental maps of the NW cross-section are shown in Figure \ref{fig:growth2}(c). For the GaAsN shell, the mean values of Ga, As, and N concentrations are 50.7, 46.4, and 2.9 atomic \% respectively (see also EDX in SI8). This N concentration in the shell is in good agreement with the PL data in Figure 1 and matches the strain analysis results obtained by geometric phase analysis (GPA), which estimates approximately 2.5\% of N, see details in SI8-SI9.

\subsection{Excitonic emission at low temperature}
\begin{figure}[] 
	\includegraphics[scale=1]{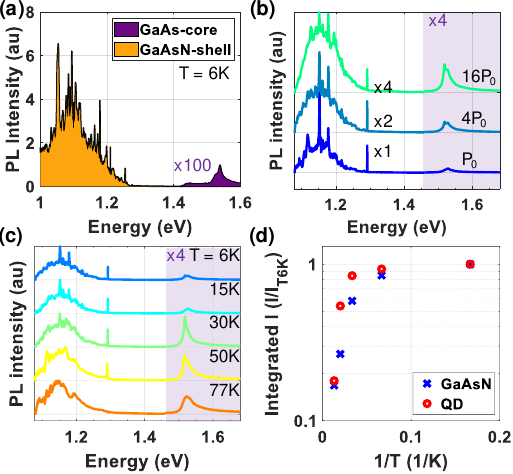} 
	\caption{Optical properties at low temperature. (a) $\upmu$-PL spectrum from the growth chip of sample B at T = 6 K. The emission from the 10 nm thick GaAsN shell is colored orange, while the emission from the 40 nm thick GaAs core is colored purple and multiplied by a factor of 100 for better visibility. (b) shows the power and (c) the temperature study of a different point on sample B. They are measured with the CCD detector, which cuts off the emitted PL signal at the low energy side of the GaAsN band. (d) shows the thermal quenching of the quantum dot emission at 1.29 eV with respect to the thermal quenching of the GaAsN shell emission in an Arrhenius plot.}
	\label{fig:LT} 
\end{figure}
First, the optical properties of the NWs are investigated directly on the NW ensemble on the Si growth chip. Figure \ref{fig:LT}(a) shows the $\upmu$-PL spectrum at 6 K from the very thin NWs of sample B with a GaAsN shell thickness of 10 nm. The spectrum has two emission bands, the low energy one from the GaAsN shell and the high energy one from the GaAs core, which was multiplied by a factor x100 to become visible. The spectral position of the GaAsN shell emission at 1.09 eV remains constant across different points on the sample, suggesting a uniform N composition across different NWs, see SI10 for further spectra from samples A and B. The GaAs core emission is more point-dependent and centered between 1.53-1.57 eV for sample A with the thinner core diameter of 20 nm and 1.51-1.53 eV for sample B with a core diameter of 40 nm. These are relatively high energies compared to the GaAs bulk of 1.515 eV or GaAs NWs with short politypic regions with emission bands between 1.51-1.52 eV \cite{DeLuca2013b}. As discussed in the SI10, the shift to higher energy for thinner GaAs core diameters is mainly affected by two mechanisms, one related to a high thermalization temperature of charge carriers, previously observed in GaAs NWs \cite{Tedeschi2016}, and the other one related to a compressive strain in the core, due to the lattice mismatch between the constituents \cite{Balaghi2019, Glas2006}. Emission peaks around 1.57 eV have been measured before on polytypic tips of GaAs NWs \cite{Somaschini2013}.

The GaAsN shell emission dominates the spectrum in all points of samples A and B, its emission intensity is up to three orders of magnitude higher than the core intensity. This is due to an efficient exciton transfer from the high-energy GaAs core to the low-energy GaAsN shell \cite{Chen2014, DeLuca2013}. 
The power and temperature study of a typical point on sample B are shown in Figures \ref{fig:LT}(b) and (c). The power study highlights that the narrow lines remain present when increasing the power by an order of magnitude before reaching the damaging threshold of the NW. The total PL emission decreases by an order of magnitude when heating the sample to a temperature of 80 K.
The weight of the high-energy side of the GaAsN emission is progressively decreased as temperature increases, as trapped carriers are thermally activated into the delocalized states, where they move to lower energy states or non-radiative recombination centers.
The pronounced sharp peaks on the low-energy emission band are typical for the GaAsN material. They are created by short-range N-concentration fluctuations leading to a three dimensional confinement of the carriers in quantum dots, as observed in bulk GaAsN \cite{Mintairov2004} with N concentration of 3\% and in thick GaAs/GaAsN NWs with low N concentration of 0.5\% \cite{Chen2014, Jansson2018, Filippov2016}, for which no single photon emission was reported. In bulk, single photon emission has been observed from luminescent centers that are localized on impurities or N-complexes. Typically, they emit at well-defined emission energies in the visible between 1.48 and 1.51 eV and dominate in very dilute N concentrations up to 0.3\% \cite{Polimeni2008, Ikezawa2017, Schwabe1985}.
The highly localized states in our NW samples have different emission energies distributed over the GaAsN band. They have a typical linewidth between 300-700 $\upmu$eV. The temperature study shows that the excitons are trapped up to a temperature of 50 K when the narrow lines disappear. The Arrhenius plot in Figure \ref{fig:LT}(d) shows the integrated intensity of the GaAsN emission band and the QD at 1.29 eV visible in (c). It highlights the better thermal stability of an exciton in a quantum dot, compared to the localized excitons in the GaAsN shell.
Finally, cathodoluminescence measurements (shown in SI11) clearly confirm that the emission from both the GaAsN shell and the GaAs core is from single vertical NWs connected to their growth substrate and not from interstitial layers.

\subsection{Single photon emission in the GaAsN NW shell}

\begin{figure}
	\centering 
	\includegraphics[scale=1]{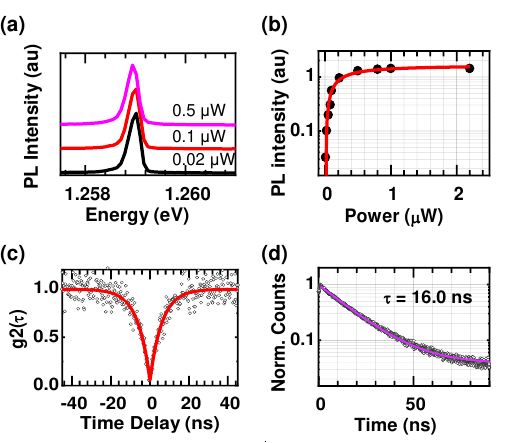} 
	\caption{Single photon emission by a strongly localized exciton in the GaAsN shell. (a) Three spectra from a power study of a narrow, spectrally isolated peak from a point close to the one of the spectrum shown in Figure 4. The linewidth broadens for increasing laser power. (b) shows the integrated area of the sharp emission peak as a function of excitation power, saturating like a typical two-level system. (c) Second-order autocorrelation measurement of the line in (a). The normalized coincidence counts are shown as a function of the time delay $t$. Single-photon emission is confirmed by a value of $g^{2}\left(0\right)$ equal to $0.056\pm 0.027$. (d) shows the time-resolved PL measurement from which a decay time of $\tau=16$ ns is extracted.}
	\label{fig:SPE} 
\end{figure}

To prove that the localization of carriers in these highly confined structures results in the emission of pure single photons, we measure the $\textrm{g}^{(2)}(\tau)$ of one of the spectrally isolated lines on the high-energy side of the GaAsN emission band of sample B. The spectral features of the line were monitored for increasing values of the excitation power. As shown in Figure \ref{fig:SPE}(a), the increasing power leads to slight redshift and linewidth broadening, due to laser-induced heating and Varshni-like shift of the bandgap. Figure \ref{fig:SPE}(b) shows the integrated area of the investigated line for increasing excitation levels, featuring a saturation behavior typical of quantum emitters and accurately described by the fitting function $I = I_{\mathrm{sat}}\left[P/\left(P+P_{\mathrm{N}} \right)\right]$ (where $I_{\mathrm{sat}}$ is the saturation intensity, $P$ is the laser power and P$_{\mathrm{N}}$ is the laser power for which the intensity is half of the saturation value). To assess the purity of the state as a single-photon emitter, $\textrm{g}^{(2)}(\tau)$ was measured with a Hanbury Brown and Twiss setup at $P=0.5 \upmu$W, which is below saturation of the quantum dot \cite{https://doi.org/10.1002/adfm.202315936}. Figure \ref{fig:SPE}(c) shows a pronounced antibunching at zero-time delay, with a value of $\textrm{g}^{(2)}\left(0\right)=0.056$, well below the threshold of 0.5, for which a state emits only one photon per cycle of excitation. The antibunching dip at $\textrm{g}^{(2)}\left(0\right)$ returns to the asymptotic value of 1 for time delays of several ns. These slow recombination dynamics are investigated with time-resolved PL measurements. The exponential decay of the emitter is shown in Figure \ref{fig:SPE}(d) and shows a decay time of 16.0 ns. 

The small values for $\textrm{g}^{(2)}(0)$ demonstrate a high purity for single photon emission in a system without any performance-enhancing cavity and underlines the high quality of the material. Its lifetime, $\tau=16$ ns, is long if compared to the decay times of QD-based single photon sources in NWs, micropillars or photonic cavities, ranging between 0.08-2.4 ns \cite{Senellart2017}. Slow exponential decay times of localized excitons in GaAsN have been observed before, with a characteristic time of 10 ns in epilayers \cite{Buyanova1999a} and 5.2 ns in NWs \cite{Chen2015}. QDs in GaAsN epilayers that are engineered by site-selective hydrogenation have shorter decay times of 1-3 ns \cite{Birindelli2014}. A range of decay times has also been observed in excitons bound to N-complexes in GaAs from approximately 0.8 ns \cite{Ikezawa2012} to more than 10 ns \cite{Schwabe1985}. While thermally activated non-radiative processes are negligible at 5 K, material defects in the emitter's surroundings may still accelerate transitions, leading to shorter lifetimes. The slow recombination rate of our emitter is a signature of its high crystalline purity and of its strong localization in the potential dips of the conduction band, which reduces the probability of trapping by fast non-radiative centers. However, the observed long decay time may also indicate that the e-h pairs are localized by spatially separated dips in the potential landscape of the conduction and valence band, and that their radiative recombination rate is, therefore, low due to the weak overlap of their wavefunctions. In conclusion, the observation of single photon emission in these NWs (a first for GaAsN-based wires, as we have noted) can be attributed to two of their defining characteristics: the reduced thickness of their GaAsN shell, which further promotes three-dimensional carrier confinement in addition to the localizing role of the N atoms, and their high crystalline purity, which had not been achieved in GaAsN NWs before.

\section{Conclusion}
In this work we have demonstrated the growth of thin GaAs/GaAsN/GaAs core multi-shell NWs via plasma-assisted MBE on Si, which makes the devices suitable for future quantum photonic circuit applications. The structural properties were analyzed by atomic resolution TEM and ultramicrotomy sectionning of the NWs. We achieved a high crystalline quality with a defect-free ZB phase along the main part of the NW and a short WZ segment at the top of each NW. The low temperature $\upmu$-PL spectra show very narrow excitonic lines from quantum dot-like states that thermalize around 50 K. This study shows the importance of the low thickness of the GaAsN shell in order to create pure single photon emitters. For the first time, we measure quantum light emission with a $g^{2}(\tau)$ value at zero time delay of 0.056 from GaAsN-based NWs. In conclusion, we have designed a NW system that is suitable to create fiber-coupled plug-and-play devices based on single photon emitters monolithically integrated on Si.

\section{Experimental}
\subsection{Growth}
The NW samples were grown by molecular beam epitaxy (MBE) in a Riber 32P system equipped with Ga and As effusion cells and a radio frequency (RF) plasma source fed by a mixture of ultra pure N$_2$ and Ar gases. 
Si (111) As-doped wafers were utilized for all samples. Samples A and B were grown on the thin native oxide layer.
NWs were grown by combining Ga-assisted VLS growth for the GaAs core and VS epitaxial growth for the GaAsN and GaAs shells. The overview of the growth procedure is depicted in Figure 1(a). In situ surface modification procedure (SMP) \cite{Tauchnitz2017} was employed for samples A and B to ensure the formation of homogeneous Ga nanoparticles (NPs) assisting the NW core growth. The procedure was as follows: first annealing at 730°C for 30 minutes, deposition of 3 monolayers (MLs) of Ga at 600°C, and second annealing at 730°C for 5 minutes. At the end of NW growth, Ga droplets were crystallized in GaAs under As flux \cite{Priante2013}. The substrate temperature was then lowered to 500°C for GaAsN and GaAs shells for all samples. The growth details of NW samples are summarized in SI12. XRD measurements of pseudomorphic thin film samples deposited under the same conditions as these NWs provide a nominal N concentration of 1.5\% for sample A and 0.9\% for sample B. However, the effective N incorporation may differ, as discussed in the SI12\cite{Li2016}.

\subsection{Imaging}
NWs were transferred onto a copper grid and analyzed using Thermo Fisher Scientific TEMs, Tecnai Osiris and Talos F200S, operated at 200 kV. Ultramicrotome cuts are created by the Leica EM UC7 Ultramicrotome system. The structures were embedded in epoxy resin and peeled out from the substrate. A Diatome ultra 35° diamond knife was used to obtain smooth cross-sections, with the thickness of each cross-section aimed to be 80 nm. Atomic-resolution images of the NWs cross-section were obtained using an aberration-corrected FEI Titan Themis STEM operated at 300 kV; more information about the EDX and GPA methods can be found in the SI.

\subsection{Optical measurements}
All the $\upmu$-PL measurements were performed using a 532nm solid-state laser (DPSS) with controlled excitation power to limit the heating and damage of the NWs. The light was focused through a 100x objective with NA=0.75, resulting in a diffraction-limited spot size of 750nm. The signal was collected in the backscattering geometry through a 0.75 NA microscope objective, dispersed by a 0.5m long spectrometer, and detected by a liquid nitrogen cooled CCD and InGaAs detector. All spectra were normalized for the response of the setup. The pumping power is 10 $\mu$W for room temperature measurements and 4 $\mu$W for measurements taken at 6K. The spot size captures emission from approximately 8 NWs.

For the  $\textrm{g}^{2}(t)$ and the time-resolved $\upmu$-PL measurements, the sample was placed on a piezoelectric stage in a closed-cycle He cryostat at a temperature of 5.6 K. For the  $\textrm{g}^{2}(t)$, the exit slit of the spectrometer was set to limit the spectral bandwidth to 1.5 nm, centered on the quantum emitter (QE) line. For measuring the coincidence counts, the dispersed signal was collimated by a parabolic mirror into a Hanbury-Brown and Twiss setup. A 50/50 beam splitter divided the light into two branches that were collected by two Si avalanche photodiodes  that were interfaced with a PicoHarp 300 time-correlated single-photon counting (TCSPC) module (maximum resolution of 4 ps). Cross-talk-induced bunching peaks were eliminated by comparing them to the coincidence counts of the $\upmu$-PL emitted from an InP sample without single-photon behavior.

For the time-resolved $\upmu$-PL measurement, we used a supercontinuum pulsed laser with a 50 ps pulse width and a maximum repetition rate of 77.8 MHz, tuned at 520 nm with acoustic-optic tunable filters. A beam sampler on the excitation path provided the START signal for measuring the time-difference at a photodiode interfaced with the TCSPC module, and the $\upmu$-PL signal from the sample is focused onto a single avalanche photodiode, providing the STOP signal. The decay curve of the emitter yields a lifetime which is considerably longer than the instrument response function (420 ps for our setup), so no deconvolution procedure had to be performed.

\begin{acknowledgement}
N.D. and M.D.L. acknowledge funding from the Swiss National Science Foundation Ambizione Grant (Grant Number $PZ00P2179801$). Authors from EPFL acknowledge funding through the NCCR QSIT. T.N. and S. R. acknownledge funding from the Horizon 2020 program of the European Union for research and innovation, under Grant Agreement No. 722176 (INDEED).
The project was funded by the European Union (ERC starting grant, NANOWHYR, 101042349). Views and opinions expressed are those of the author(s) only and do not necessarily reflect those of the European Union or the European Research Council Executive Agency. Neither the European Union nor the granting authority can be held responsible for them.
A.P., M.F. and M.D.L. acknowledge funding from the PNRR MUR project PE0000023-NQSTI (the National Quantum Science and Technology Institute). The authors acknowledge Prof. Ilaria Zardo and Dr. Elena Blundo for experimental contribution and fruitful discussions.
\end{acknowledgement}

\begin{suppinfo}

\begin{itemize}
  \item Supplementary information: Structural characterization of samples A and B, HR-TEM of nanowire tips and of bottom parts, EDX of sample A, strain analysis, cathodoluminescence, additional photoluminescence measurements.
\end{itemize}

\end{suppinfo}

\section{Author contributions}
T.N. and S.R. performed NW growth and SEM characterization. D.D., V.B. and A.F.M. performed and analyzed TEM measurements of NWs, ADF-STEM, EDX and cathodoluminescence. S.C., M.D.L, M.F. and A.P. performed the  $\textrm{g}^{2}(t)$ and the time-resolved $\upmu$-PL measurements at low temperature. N.D., F.S. and M.D.L performed room temperature $\upmu$-PL measurements. N.D. and M.D.L. performed temperature-dependent, power-dependent, and point-dependent $\upmu$-PL measurements. N. D. analyzed all $\upmu$-PL data and wrote the manuscript, with contributions from all coauthors. M.D.L. conceived the project.

\bibliography{ThinWires.bib}	

\end{document}